# FLYEYE FAMILY TREE, FROM SMART FAST CAMERAS TO MEZZOCIELO

R. Ragazzoni[1,2,6], S. Di Rosa[1], C. Arcidiacono[2], M. Dima[2], D. Magrin[2], A. J. Corso[3], J. Farinato[2], M. Pelizzo[4], G. L. Santi[5], M. Simioni[2], S. Zaggia[2]

[1] Univ. of Padova, Dept. of Phys. & Astron., vic. Osservatorio 3, I-35122 Padova (Italy)
[2] INAF – Astron. Obs. of Padova, vic. Osservatorio 5, I-35122 Padova (Italy)
[3] CNR, Istituto di Fotonica e Nanotecnologie, via Trasea 7, 35131 Padova (Italy)
[4] Univ. of Padova, Dept. of Information Engineering, via Gradenigo 6B, 35131 Padova (Italy)
[5] Univ. of Padova, Centro di Ateneo Studi e Attività Spaziali, via Venezia 15, 35131 Padova (Italy)
[3]Email: roberto.ragazzoni@inaf.it

**ABSTRACT**

We developed game-changing concepts for meter(s) class very-wide-field telescopes, spanning three orders of magnitude of the covered field of view. Multiple cameras and monocentric systems: from the Smart Fast Cameras (with a quasi-monocentric aperture), through the FlyEye, toward a MezzoCielo concept (both with a truly monocentric aperture). Mezzocielo (or "half of the sky") is the last developed concept for a new class of telescopes. Such a concept is based on a fully spherical optical surface filled with a low refractive index, and high transparency liquid surrounded by multiple identical cameras. MezzoCielo is capable to reach field of views in the range of ten to twenty thousand square degrees.

## 1 PROLOGUE

Multiplexing is a key issue in astronomical instrumentation. This is somehow different from other experimental sciences, like particle physics, where a very powerful single machine can convey all the possible experiments doable with that specific power, leaving little room for duplication, for instance, of the same kind of accelerator. In contrast, as a purely observative science, astronomy benefits from the multiple observations of the same class of objects, leading, in the most simples form, to the need to cover a very wide Field of View (FoV) of the sky in order to investigate a variety of transient and unexpected phenomenon. It is noticeable that this feature coincides with the patrol of the sky for NEOs and space debris. In the beginning of this century, I had the opportunity to translate into reality the wide FoV cameras conceived for large telescope: namely a couple of Prime Focus correctors [1,2,3,4] now installed at the foci of the two 8.4m parabolic mirrors of the Large Binocular Telescope [5]. This adventure ventured me and a small group of astronomical instrumentalists into handling heavy and large (about 80kg and 820mm in diameter) lenses built by fused silica (a rather fragile kind of glass) and BK7. We built two of these units optimized respectively for the bluer portion of the electromagnetic spectrum that still reach the ground and for the redder one with provision for the near InfraRed one (although this option has not yet been exploited).

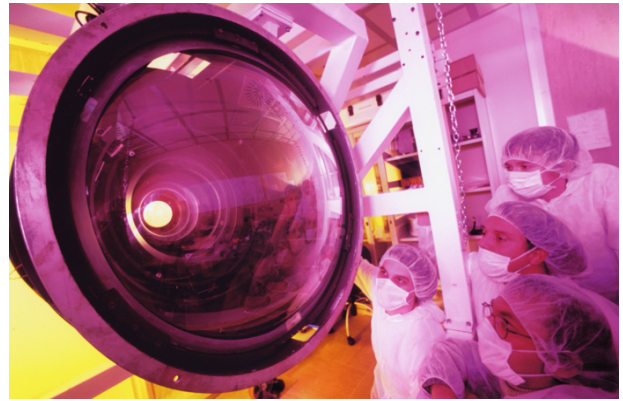

*Figure 1: The Blue channel of the LBT Prime Focus during the final integration phase in the lab. The main lens is an impressive 800mm class fused silica meniscus. The first author is depicted in the lower right side of the picture.*

Although later similar instrumentation has been built [6] our couple of Prime Focus is unique on this class of telescopes for their very high efficiency in the U-band down to the wavelength of 350nm (as, for instance, on the smaller class 3.6 meter Canada France Hawaii u-band [7]) and, for the last 15 years, has remained the most productive instrument - in terms of published papers - onboard the telescope built by an US-Italian and German consortium. This, of course, marked an history of success, however the efforts profuse and the difficulties encountered have been impressive. Most of them, as usually happens, have not been predicted in advance. Indeed, although prime focuses are supposed to be - from several point of view - a kind of conventional instrument, in our specific case we pushed the limits toward the ultraviolet coverage and, thanks to the rather fast (F/1.14) foci of the primary mirrors, designed at an epoch in which no prime focal station was ever planned. Stimulated by this experience, we thought it was time to change the paradigm in the wide field astronomical instrumentation and we started developing a step beyond that could take advantage of the lessons learned. Therefore, we conceived a post telescope focal plane array of correctors built in a sort of replicated manner. Each individual unit covers a portion of the FoV small enough to assume constant off-axis aberrations over such a solid angle. The correctors were supposed to be built by a sort of aberrating plates that cancel each other if properly co-aligned. The relative rotation of these plates would translate into a changing aberration compensator unit. In this way the rotation of the ensemble of the two plates would identify the position angle of the non-symmetric portion of the aberration. This was somehow resembling systems for offsetting reference stars in telescope systems (like MAST [8] or SIMURIS [9], to position on the focal plane specific portion of the solar surface) and acts as the two main sections of a robotic arm.

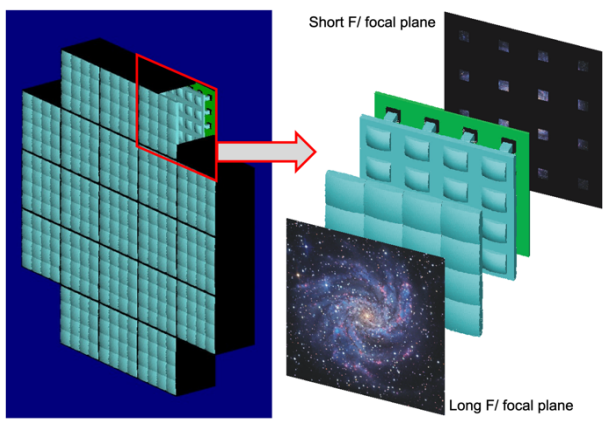

*Figure 2: The Smart Fast Camera concept employs a large number of almost identical focal reducer with an intermediate pupil plane where the - varying over the Field of View - aberration is being compensated by cleverly mounted fixed optical plates or electro-optical devices.*

Although we developed such an approach [10,11] in some details and planned a preliminary design in order to propose the construction of such an instrument engineered as a wide field spectrograph [12,13,14] to cover a significantly larger fraction of the sky at the focal plane of one of the VLTs units, we soon realized that the variety and combination of the rapidly changing aberrations across the intermediate FoV would have required more than the two plates originally conceived. In fact, much later, it was implemented taking advantage of the technological development, which, in the meantime, allowed substituting the passive plates by electro-optics unit [15,16]. Unfortunately, this solution was neither available or low cost at the time of conception of what we called pompously a "Smart Fast Camera". Further to the lack of available technology, the concept suffers just because of the varying aberrations. Removed this limitation, it would retain the advantages of the mass production of the compensating cameras.

## 2 THE FLYEYE CONCEPT

The use of a primary spherical mirror retains together a number of advantages: the lack of a defined axis of symmetry (in fact the spherical mirror is a kind of monocentric device where the only symmetry is retained around the center of curvature), the easiness of polishing, and of segmentation (all the segments retain the same optical figure, in contrast with parabolic or hyperbolic mirrors, characterizing, for instance, the entrance apertures of, respectively, Cassegrain and Ritchey-Chretienne two mirrors telescope). All this comes to a price, namely the residual spherical aberration, often completely unacceptable. The latter is, however, constant over the ubiquitous line of sight over the FoV. Usually, the compensation of such aberration is easily accomplished by an optical system made up by at least two elements: the first one forms an image of the entrance pupil onto the second one which provides the spherical compensation needed. This approach has been widely used in radio telescopes [17] so large that it would be unconceivable or too expensive to establish a system to point the whole structure toward a certain direction, or where a low-cost large aperture is being implemented for a very narrow science case in the optical domain [18].

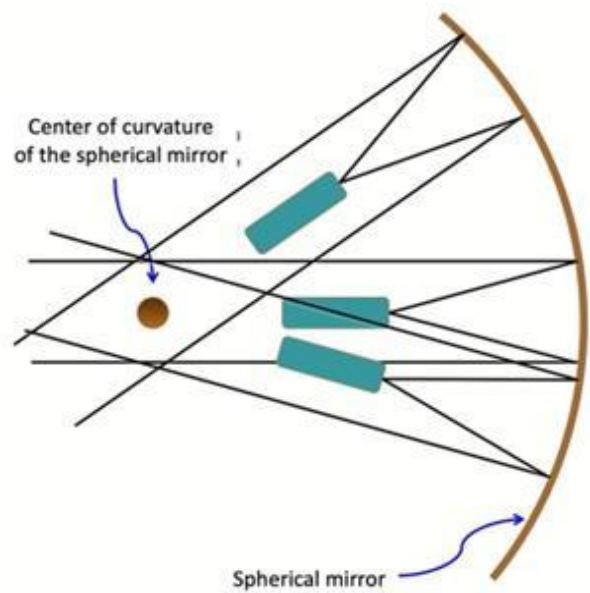

*Figure 3: The FlyEye concept relies on the use of multiple identical correctors mounted on the intermediate focal plane of a common spherical primary mirrors. The cameras are looking at the monocentric converging device through a number of folding mirrors*

*located in order to minimize as much as possible the obstruction of such an approach. This obstruction - in contrast with the Schmidt telescope - is the only limit to the Field of View achievable in this context.*

Multiplexing such a corrector has been the straightforward extension of the Smart Fast Camera concept to a common spherical primary mirror. It is noticeable that field lenses are located somehow after the intermediate (and heavily aberrated) focus of the spherical main mirror, where a number of folding mirrors conveniently redirect the beams to a small array of cameras. The FlyEye concept ([19,20,21,22,23,24,25,26,27]) turned slowly from conceptual drawings to blueprints and real prototypes.

## 3   SCHMIDT CONFRONTATION

The obvious competitor for this kind of approach is the classical Schmidt-like telescope. In this approach, the spherical aberration is corrected for all the employed line of sights with a single refractive corrector located on the center of curvature of the telescope. Several variations on the theme are possible, including UV-compliant where the spherical aberration should be a reflective device. While a number of practical different approaches are beyond the limit of this short manuscript, one should take note that there is a conceptual difference between the Schmidt and the FlyEye approach. The Schmidt plate, in fact, retain an axis of symmetry, although - as a refractive device - the deviations off-axis are initially mild.

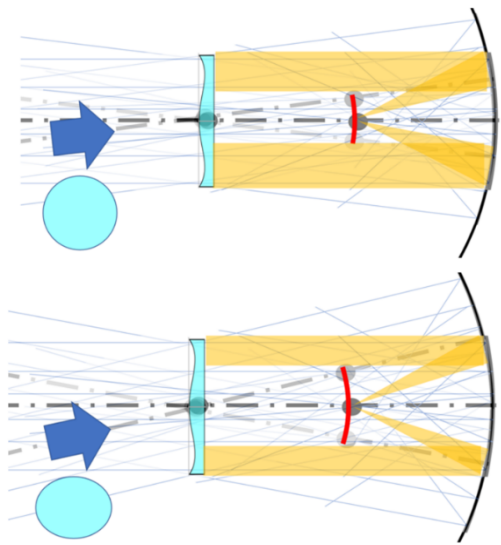

*Figure 4: In a conventional Schmidt telescope there are two limits for the achievable Field of View. One is the central obstruction, depicted in red in these two panels of moderate (upper) and large (lower) off-axis acceptance angle. The other is being represented by the sub-optimal compensation of the spherical aberration.*

*The Schmidt plate, in fact, is progressively seen in an obliqued manner, leading to a sub-optimal compensation of the spherical, centrosymmetric, aberration.*

However, the larger the off-axis angle, the less effective is the compensation. In contrast, the Field of View of the FlyEye is only limited by the obstruction introduced by the optical elements close to the intermediate focus of the spherical mirror. Namely the folding mirrors and -depending upon the details of the optical design- the field lenses and, of course, their mounts. In the Schmidt telescope, even without taking account the vignetting issue caused by the focal plane detector, at a certain off-axis angle the quality and efficiency of the correction plates would become so deteriorated to make the image quality unacceptable.

## 4   MONOCENTRICITY WITHOUT OBSTRUCTION: MEZZOCIELO

The FlyEye concept is, in a certain sense, revolutionary, as there is no conceptual limit to the Field of View. The practical limit, however, is rather stringent as in practice both the folding mirrors and the field lenses are somehow inserted and obstructing the entrance beam. Furthermore, if you want a continuous uninterrupted large patch in the sky properly reimaged onto an array of detectors, then you are forced to place additional folding mirrors much out of focus than what can be achievable with a smaller FoV. Despite this, FlyEyes are able to easily surpass Schmidt telescopes in terms of practically achievable FoV but, somehow frustratingly, are unable to exploit in full the inherent conceptually unlimited field potential. This is overcome using a monocentric refractive solution. It is not necessarily a new approach, but, in order to keep the spherical aberration controllable, the monocentric refractive system doable, and the transparency high enough not to be limited by such a parameter, we conceived a solution where a spherical hollow optical system is filled with industrial liquids with low refractive index and extremely high transparency. It is noticeable that such fluid turned out to already exist and used for non-optical applications. As, in principle, this solution can achieve the continuous observations of the whole sky available at any given location on Earth, we nicknamed it "MezzoCielo" (Italian name for "half of the sky", [28,29,30,31,32,])

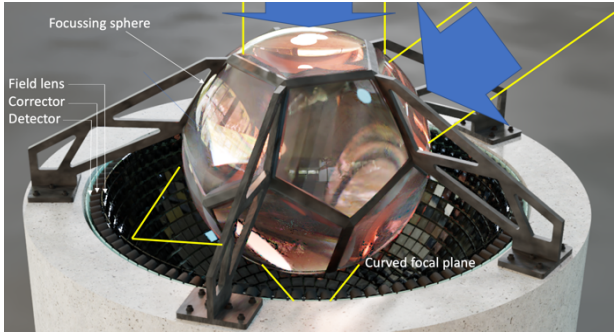

*Figure 5: In the MezzoCielo concept a number of meniscus lenses forms a hollow sphere filled with high transparency fluids. Using lenses of the same class used in the beginning of the story depicted in this manuscript, a sphere of diameter comparable to the effective aperture of the current FlyEye prototypes is achievable. However, the covered Field of View is more than two orders of magnitude larger.*

This concept (whose patenting has been very recently approved by the European Patent Office) is a game changer from several viewpoints. First of all, its cost is dominated by the focal plane detectors and cameras, although they are all identical and hence are expected to be built with a mass production approach. The use of CCDs would make this extremely expensive (although arrays of identical cameras with the same sky coverage and pixel size would lead to an identical cost with a much smaller effective area), hence the emergent CMOS detector technology is probably the most viable solution. Furthermore, as the telescope is nominally looking at the whole sky, pointing becomes a useless capability. Long exposures, if needed, could still benefit from a sort of equatorial movement of the solely camera system. Patrolling of transient phenomenon is probably not needing even this residual degree of freedom, making this class of telescopes free of any necessity of mounting. Of course, a number of issues are to be investigated, including the requirements in terms of optical precision of the meniscus that could led, in turn, to the requirement of an active control of the solid elements. Even if the system would work in seeing limited mode, in fact, the coaddition of beams coming from different meniscus would translates into a plate scale precision requirement that would make the optics with tolerances comparable to a diffraction limited system, unless to implement more sophisticated data handling approaches.

## 5   A PROPOSAL FOR AN EPILOGUE

We are currently preparing a proposal to build a half a meter class prototype and a cylindrical section of a full one meter class "MezzoCielo", in order to get enough experience and hand on feedback to proceed to construct a wide field telescope that would outperform most of the existing all sky surveys. Probably, with the sole exception of what can be achieved, on a much longer time frequency domain, by the Vera Rubin's telescope, patrolling Near Earth Objects (NEO) and space debris would benefit of a two orders of magnitude larger instantaneous covered FoV and with a much higher Signal to Noise Ratio (SNR) for tracklets. These could be tracked for a much longer time. This approach can give some gain only if the instantaneous SNR is larger than a given amount, probably making this approach un-useful for the very smaller sources. This could be compensated by larger monocentric systems. With the current fluid we are examining, that was not engineered on purpose, the size at which the transparency of the fluid made a larger sphere no more effective, can fall at diameter that makes 4m class telescopes conceivable. This will come with a number of additional technical difficulties that are not even listed here. If the Moore's law does apply on detectors and data handling system as well, it is not difficult to forecast that this approach is the one that has a vivid future in the short to medium timescale. The diffidence that the previous conceived solutions encountered should be a lesson from which to learn to invest major efforts into the further development of the MezzoCielo concept.